\documentclass[preprint2]{aastex}

\usepackage{graphicx}

\newcommand{\beq}{\begin{equation}}
\newcommand{\eeq}{\end{equation}}
\newcommand{\beqa}{\begin{eqnarray}}
\newcommand{\eeqa}{\end{eqnarray}}

\begin{document}
\title{Massive Coronae of Galaxies}
\author{Masataka Fukugita}
\affil{Institute for Cosmic Ray Research, University of Tokyo,\\
            Kashiwa 277-8582, Japan}
\and\affil{Institute for Advanced Study, Princeton, NJ 08540 USA}
\author  {P. J. E. Peebles}
\affil{Joseph Henry Laboratories,
Princeton University,\\
Princeton, NJ 08544 USA}
\begin{abstract}
There is reason to suspect that about half of the baryons are in pressure-supported plasma in the halos of normal galaxies, drawn in by gravity along with about half of the dark matter. To be consistent with the observations this baryonic component, the  corona, would have to be hotter than the kinetic temperature of the dark matter in the halo so as to produce acceptable central electron densities. We ascribe this hotter plasma temperature to the addition of entropy prior to and during assembly of the system, in an analogy to cluster formation. The plasma cooling time would be longer than the gravitational collapse time but, in the inner parts, shorter than the Hubble time, making the corona thermally unstable to the formation of a cloudy structure that may be in line with what is indicated by quasar absorption line systems. The corona of an isolated spiral galaxy would be a source of soft X-ray and recombination radiation, adding to the more commonly discussed effects of stars and supernovae. In this picture the mass in the corona  is much larger than the mass in condensed baryons in a spiral galaxy. The corona thus would be a substantial reservoir of diffuse baryons that are settling and adding to the mass in interstellar matter and stars, so that star formation in isolated spirals will continue well beyond the present epoch. 
\end{abstract}

\section{Introduction}

Dynamics and weak lensing suggest that a significant fraction  of the dark matter is in the virialized halos of $L\sim L_\ast$ galaxies. Gravity would have tended to draw a like fraction of the barons into the galaxies. Since the observed mass in condensed baryons --- stars and the interstellar medium --- is only about 8\%\ of the total we have been led to consider the idea  that roughly half of the baryons are in pressure-supported plasma in the outer parts of the galaxies (Fukugita, Hogan \& Peebles 1998; Fukugita \& Peebles 2004, FP). This is a modern variant of Spitzer's (1956) galactic corona, and the extent of the coronae we are considering, about 300~kpc radius, is in line with the proposal by Bahcall \& Spitzer (1969) that galactic  coronae may be responsible for quasar absorption lines. The massive corona picture is related to the evidence that X-ray detected groups contain cosmologically appreciable mass in plasma (Mulchaey 2000), and the search for a possibly significant plasma mass in the Local Group (Rasmussen \& Pedersen 2001 and references therein). Since groups are not much larger than the virialized parts of an individual galaxy the constraints on the plasma properties are similar. Here we are considering the possibility that there is a cosmologically significant mass in the coronae of individual $L\sim L_\ast$ spiral galaxies, and our goal is to present an observationally acceptable model for the present-day structure of such a massive corona.

The starting idea for the model is that the bulk of the baryonic mass of a galaxy is in plasma gravitationally bound to the galaxy at near hydrostatic pressure equilibrium and at density low enough that the cooling time is longer than the Hubble time. At smaller radii the shorter cooling time would cause some of the plasma to settle toward the center of the galaxy, tending to relax to a near steady flow, in analogy to a cooling flow of intracluster plasma (e.g. Fabian 1988; Sarazin 1988).  The first point to consider is that the plasma cooling time for most of the baryons in a virialized halo can be longer than the Hubble time, meaning the massive coronae can still be present if initial conditions allow the needed minimal clumping of the plasma. This is discussed in \S 2.  

Our model for the halo mass distribution to which the corona is gravitationally bound differs from FP in two ways. First, we adjust the velocity dispersion to the value appropriate for an $L\sim L_\ast$ spiral galaxy. Second, we allow for the possibility of a less sharp truncation of the edge of the halo mass distribution, which we show can produce a more reasonable-looking galaxy-mass cross correlation function. We allow the possibility that heating of the corona as it collapsed made it warmer than the temperature characteristic of the dark matter, in analogy to what is observed in clusters of galaxies (Allen et al. 2004). We assume that in the inner parts of the corona, where the cooling time is less than the Hubble time, the plasma  typically has relaxed to a near steady state, not seriously disturbed by recent merging events or by energy sources such as supernovae, massive stars, or an active galactic nucleus.  Since quasar absorption lines show quite modest mass in atomic hydrogen at radii $\ga 30$~kpc around galaxies (Chen et al. 2001), our model requires that the rate of conversion of plasma to condensed forms of baryons is significant only near the optically luminous parts of the galaxy. The model should also meet the requirement that the massive halo is consistent with soft X-ray
observations of nearby galaxies and the integrated background.
We construct two examples of corona models under these conditions  on the simple plan of spherically symmetric quasi-isothermal structures. Section 3 explains the models and compares them to the observations. We discuss possible directions for further work in \S 4.

There are alternative pictures for the present state of the baryons originally associated with the dark matter now in halos of galaxies. Shocks may have made most of this baryonic matter too hot to fall into massive halos. This is in line with the idea currently under discussion that there is a substantial baryon mass in warm-hot intergalactic matter at temperatures in the range $\sim 10^5 - 10^7$~K (e.g. Dav\'e et al. 2001), but it is difficult to see what could have produced a local shock powerful enough to have prevented ongoing accretion of baryons by the Milky Way and Andromeda galaxy: the Virgo Cluster seems to be too far away, the galaxy velocity dispersion near the local sheet of galaxies seems to be too small, and if gravity produced the local void it would not have been accompanied by production of a shock. Other ideas assume that the baryons did fall into halos along with the dark matter. Possible, though we suspect unlikely, is the idea that 
  most of these baryons collapsed to a condensed form that is not readily observable, such as brown dwarfs. These condensed baryons would have to be distributed through the dark halo, because there is not room for them in the luminous parts of normal galaxies. This could be indicated by the extended diffuse glow of light around M87 (Arp and Bertola 1969), but it requires a considerable  departure from observed initial stellar mass functions; it may be related to the local MACHO population (Alcock et al. 2000), though the indicated mass range, between $0.15$ and 0.9~$M_\odot$ if a halo component, is a problem. Silk (2003) develops the idea that supernova-driven winds may have blown most of the baryons out of the massive halos of protogalaxies. The energy required is modest but, as Silk discusses, the efficiency of entrainment might be  questioned. A low density wind would tend to drive convection through the coronae, leaving behind most of the baryons,  while locally deposited energy likely is radiated away rather than being adding to the entropy of the ambient material. Our conclusion is that the alternative, the massive corona picture, is worth exploring. 

\section{The Cooling Rate in the Outer Corona}

This discussion of the characteristic plasma cooling time for baryons in a virialized halo takes all parameters from FP. It is not significantly affected by the adjustments to the halo mass model discussed in the next section. 

We start with a measure of the plasma density at the nominal virial radius $r_{200}$ of a newly relaxed halo. The density of mass contained by $r_{200}$ is 200 times the critical Einstein-de Sitter value, and the mass density $\rho_{200}$ at $r_{200}$ is one third of that if the density run within $r_{200}$ is $\rho\sim r^{-2}$. To simplify this discussion we take it that the baryons and dark matter have the same spatial distribution, so the mass fraction in baryons at $r_{200}$ is the ratio of the mass density parameters $\Omega_b=0.045$ and $\Omega_m=0.28$ in baryons and in all matter. (The baryon segregation discussed in \S 3.3 slightly reduces the cooling time at $r_{200}$.) With primeval helium abundance $Y_p=0.248$ the mean mass per particle is $\mu = 0.59$ times the proton mass $m_p$, and this measure of the plasma particle number density at $r_{200}$ is
\beq
n_{200}={\Omega_b\rho_{200}\over\Omega_m\mu m_p} =  1.0\times 10^{-4}\hbox{ cm}^{-3},
\label{eq:virial_number_density} 
\eeq
with electron density $n_{e,200}=0.52n_{200}$.
The one-dimensional mass-weighted velocity dispersion averaged over all galaxy types is $\sigma=160$ km~s$^{-1}$, and the effective temperature is $T_v = \mu m_p\sigma ^2/k= 1.8\times 10^6$~K. At plasma temperature $T_v$ and our nominal heavy element abundance, $Z=10^{-1.5}$ times Solar, the Sutherland \&\ Dopita\footnote{Here and in the rest of the paper we use the Sutherland \&\ Dopita (1993) values for the mean mass per particle and the ratio of the electron to total particle number densities. We do not attempt to adjust the primeval helium abundance to the larger present estimate because the correction is small and the adjustment likely to be confusing. We use the corrected energy density in http://www.mso.anu.edu.au/~ralph/.} (1993) cooling function is $\Lambda = 10^{-23.01}$ erg~cm$^{3}$~s$^{-1}$, and the characteristic cooling time at $r_{200}$ is
\beq
\tau_{200} = {1.5 n_{200} k T_v\over \Lambda n_{e,200}(n_{200} - n_{e,200})} = 50 \hbox{ Gyr}.
\label{eq:tauv}
\eeq
If the velocity dispersion $\sigma$ is independent of time the plasma density at $r_{200}$ in this expression scales as the square of Hubble's constant and the ratio of cooling to Hubble times in a smooth corona varies with redshift as 
\beq
H\tau_{200} = 3.5[\Omega_m(1+z)^3 + 1 -\Omega_m]^{-1/2}.
\label{eq:Htau}
\eeq

Quasar heavy element absorption lines at projected distances $\la100$~kpc from galaxies (Churchill et al. 2000 and references therein) show that the coronae would have to have been enriched in heavy elements, but the abundance could be well below our nominal value. At primeval abundances the cooling time is twice equation~(\ref{eq:tauv}).

We conclude that the plasma cooling time near the virial radius of a present-day galaxy can be longer than the Hubble time, assuming what seems to be a reasonable heavy element abundance. This is what allows the development of a model for a massive corona that has not yet significantly  settled to condensed forms of baryons. The corona would have to have started accumulating at relatively low redshift. Equation~(\ref{eq:Htau}) indicates that at redshift $z=3$ the cooling time of a massive halo at the smaller virial radius that obtained then is less than the expansion time then, even for a smooth halo. Thus we expect that the baryons that had collected at $z=3$ largely collapsed to condensed forms within an expansion time. It is important that even at near primeval abundances the present cooling time in the outer part of a massive halo is not much greater than the Hubble time, meaning that only a modest degree of clumping of the plasma would have caused most of a massive corona to have dissipatively collapsed before the present epoch. It is reasonable therefore that in some simulations 25 to 40\% of the corona has ended up now as condensed matter (Dav\'e et al. 2001). The puzzle is that the observed condensed mass fraction is much smaller, about 8\%\ in the FP estimate. The resolution we are considering is that the corona was assembled in a smooth enough way that the cooling time for the bulk of the baryons in the galaxy once assembled was close to equation~(\ref{eq:tauv}), assembly of the major pieces of a spiral galaxy having been largely completed at $z\ga 1$.

\section{The Corona Model}

We commence with a review of the FP conventions for the halo mass structure (Model 1) and then present  an argument for an adjustment that preserves the total mass but distributes it in a way that gives a better fit to the galaxy-mass cross correlation function (Model 2). Parameters appropriate for the massive halo of a spiral galaxy are presented in \S 3.2, the corona models are developed in \S 3.3,  and the observational tests are discussed in \S\S 3.3 and 3.4. 

\subsection{The Mass Distribution}

In the halo model in FP, which is the basis for our corona Model~1, the total mass density $\rho$ as a function of radius $r$ is
\beq  
\rho (r)  = {\sigma ^2\over 2\pi G r^2}, \qquad r\leq r_v
={\sigma\over\sqrt{50}H_0}.
\label{eq:isosph}
\eeq
The circular velocity is constant at $v_c=\sqrt{2}\sigma$ at $r<r_v$, the mass within the radius $r_v$ is
\beq 
M_v = 2\sigma^2r_v/G, 
\label{eq:Mv}
\eeq
and the mean density of mass within $r_v$ is 200 times the Einstein-de Sitter value, that is, $r_v=r_{200}$. In the model to be discussed next we keep $r_v$ at the value in equation~(\ref{eq:isosph})  but adjust the mass distribution in a way that decreases $r_{200}$. 

The FP estimate of the one-dimensional mass-weighted velocity dispersion averaged over galaxy types is  
\beq
\sigma =160\hbox{ km s}^{-1}.
\label{eq:sigmaFP}
\eeq
The effective number density of galaxies is 
\beq
n_g=0.017h^3\hbox{ Mpc}^{-3},
\label{eq:ng}
\eeq
where the Hubble parameter is $H_0=70$ km s$^{-1}$ Mpc$^{-1}$, that is, $h=0.7$.  With these numbers the mass fraction within distance $r_v$ from the center of a galaxy is
\beq
f_g = {M_vn_g\over\Omega_m\rho_{\rm crit}} =0.6.
\label{eq:fg}
\eeq 
The baryon mass fraction $1-f_g$ beyond $r_v$ might be assigned to the intergalactic medium, though a substantial part may be concentrated near galaxies. 

In our corona Model~2 we adjust the halo mass distribution to produce a less abrupt truncation at $r=r_v$ and, as will be discussed, a better approximation to the galaxy-mass cross correlation function. One might consider smoothing the truncation of $\rho(r)$ in equation~(\ref{eq:isosph})   by adding mass at $r>r_v$ with a steeper density gradient, but that would make the baryon mass in galaxies unreasonably large, as one sees from the substantial fraction of mass within $r_v$ in equation~(\ref{eq:fg}). We instead keep the effective galaxy mass fixed at $M_v$,  thus holding fixed the baryon fraction in equation~(\ref{eq:fg}), and spread out the mass distribution near $r_v$. This is implemented in the model 
\beqa 
&&\rho (r)  = {\sigma ^2\over 2\pi G r^2(1 + \eta r/r_v)}\,,\nonumber\\
&& r\leq r_x=r_v{e^\eta - 1\over\eta},
\label{eq:modisosph}
\eeqa
with $\rho(r) = 0$ at $r>r_x$. The total mass in this model agrees with $M_v$ in equation~(\ref{eq:Mv}). The radius $r_v$ here serves as an analog of the break radius $a$ in the Hernquist (1990) form or $r_s$ in the  Navarro,  Frenk,  \& White (1996) model, but the form we are using has the steeper inner slope wanted to fit spiral rotation curves and weak lensing measurements. Note that the mass density near the center, set by $\sigma$, is the same as that in Model 1. The limit $\eta\rightarrow 0$ is equation~(\ref{eq:isosph}). In our corona Model~2 we set $\eta = 1$.

\begin{figure}[t]
\plotone{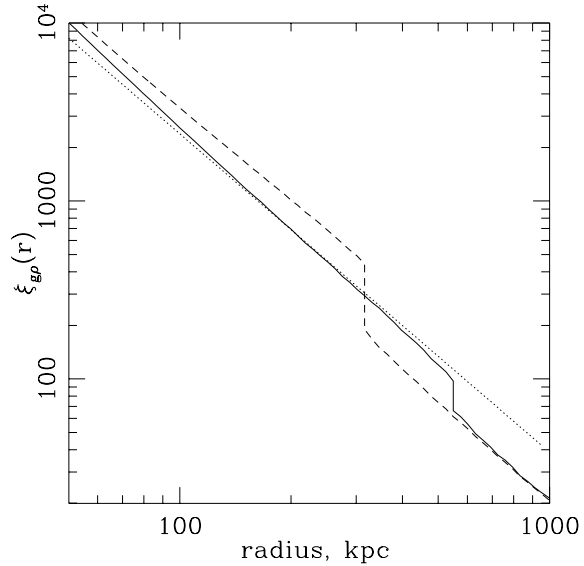}\caption{Galaxy-mass cross correlation functions. The dashed line is the $r^{-2}$ halo Model~1 in equation~(\ref{eq:isosph}),  the solid line is the halo Model~2 in equation~(\ref{eq:modisosph}) with $\eta=1$, and the dotted line is the SDSS weak lensing measurement.}\end{figure}

The model for the halo structure is tested by the galaxy-mass cross correlation function $\xi_{g\rho}$, which we approximate as (Seldner \& Peebles 1977; Cooray \& Sheth 2002)
\beqa
\xi_{g\rho}(r) = {f_g\theta(r)\over 
4\pi n_gr_vr^2(1+\eta r/r_v)}\nonumber\\
+{f_gr_o^\gamma\over 2r_v}\int 
{\theta({\rm d}){\rm d}^{2-\gamma}d{\rm d}\,d\mu\over l^2(1+\eta l/r_v)}.
\eeqa
The step function, $\theta (r)=1$ at $r<r_x$ and $\theta (r)=0$ at $r\ge r_x$, represents the extent of the halo. The first term on the right hand side is the contribution to the cross correlation function from the mass belonging to the galaxy. The second term is the convolution of the mass distributions in neighboring halos with the halo autocorrelation function, which we approximate as $\xi=(r_o/r)^\gamma$ with $hr_o=5$~Mpc and $\gamma = 1.8$. The distance to a neighboring halo is ${\rm d}$ and $l=\sqrt{{\rm d}^2+r^2-2r{\rm d}\mu}$ is the distance from the center of the neighbor to the point of observation of the mass density. This expression assumes the mass fraction $1-f_g$ (eq.~[\ref{eq:fg}]) not in halos is homogeneously distributed, which seems reasonable at $r\sim 100$~kpc but at $r\ga 1$~Mpc may underestimate $\xi_{g\rho}$. 

The dashed line in Figure~1 is the galaxy-mass cross correlation function $\xi_{g\rho}$ for the $\rho\propto r^{-2}$ halo model in equation~(\ref{eq:isosph}), the solid line is $\xi_{g\rho}$ for the halo model in equation~(\ref{eq:modisosph}), and the dotted line is the Sheldon et al. (2004) power law fit to their weak lensing measurement, $\xi_{g\rho}=(5.4\hbox{ Mpc}/hr)^{1.79}$. FP chose the $\rho\propto r^{-2}$ model to agree with the $\xi_{g\rho}$ measurement as well as the circular velocities observed at smaller radii. The fit to the former is indeed close, but $\xi_{g\rho}$ is somewhat high at $r\sim 200$~kpc because halos overlap, and the abrupt cutoff in $\rho(r)$ produces a pronounced shoulder in $\xi_{g\rho}$ at $r=r_v$. The steeper halo density run at $r\sim r_v$ in the $\eta=1$ model removes the excess and reduces the shoulder. A still larger value for $\eta$ would make the shoulder even smaller but it would produce an undesirable dip in $\xi_{g\rho}$ at $r\sim 100$~kpc. At $r>r_x$ both models fall below the measured $\xi_{g\rho}$. We attribute this to the mass fraction $1-f_g$ that we suspect is clustered on scales $\sim 1$~Mpc. 

We conclude that the models with $\eta\rightarrow 0$ and $\eta=1$ both give useful approximations to the massive halos of $L\sim L_\ast$ galaxies, within the considerable uncertainties, but that the $\eta=1$ case produces the better fit to the galaxy-mass cross correlation function. 

\subsection{The Massive Halo of a Spiral Galaxy}

We are considering the idea that field galaxies have massive coronae with baryon mass fractions close to the cosmic mean. For late galaxy types it is appropriate to use a lower velocity dispersion than the average over types in equation~(\ref{eq:sigmaFP}). We adopt the FP value for spirals, 
\beq
\sigma _{\rm late}=140\hbox{ km s}^{-1}. 
\label{eq:sigma_late}
\eeq
This defines the effective temperature,
\beq
T_v = \mu m_p\sigma_{\rm late}^2/k= 1.4\times 10^6 K,
\eeq
and the late-type halo parameters (eqs.~[\ref{eq:isosph}] and [\ref{eq:Mv}]),
\beq 
r_v = 280\hbox{ kpc}, \quad
M_v = 2.6\times 10^{12}M_\odot.
\eeq
An estimate of the mean mass of the Milky Way and the Andromeda galaxy from the dynamics of relative motions of the nearby galaxies is $2.3\times 10^{12}M_\odot$ (Peebles 1995), reasonably close to $M_v$ with our choice of $\sigma_{\rm late}$.

The conventional definition of the virialized part of a galaxy is that within the radius $r_{200}$ at which the enclosed density of mass is 200 times the Einstein-de Sitter value. In our late-type halo Model~1, with $\eta\rightarrow 0$ in equation~(\ref{eq:modisosph}), $r_{200}=r_v=280$~kpc. In Model~2, with $\eta=1$, the density of mass within $r_x$ is 40 times the Einstein-de Sitter value, or 140 times the present mean matter density. The lower density in the outer parts of this model  reduces the conventional virial radius to $r_{200}=240$~kpc and it reduces the density parameter of the mass within $r_{200}$ in spiral galaxies from $\Omega_v=0.08$ in Model~1 to $\Omega_v=0.05$ in Model~2. One should bear in mind, however, that the region of near dynamical equilibrium may extend beyond $r_{200}$.

\subsection{The Corona of a Spiral Galaxy}

 In the FP inventory the condensed mass is 13\%\ of the total  baryon mass in a galaxy (from eq. [27] and entries 3.1a, 3.9 and 3.10 in Table~1 in FP), leaving the mass
\beq
M_{\rm corona} = 0.87M_v\Omega_b/\Omega_m=3.6\times 10^{11} M_\odot
\label{eq:M_corona}
\eeq
for the massive corona of our characteristic spiral galaxy.

Since the cooling time in the denser inner part of this corona is shorter than the Hubble time, and the Milky Way seems not to have been substantially disturbed by mergers since redshift $z\ga 1$ (Gilmore, Wyse \&\ Norris 2001), we are assuming that the inner corona of a spiral typically has had time since the last major merger to have relaxed to a near steady rate of settling. This steady state means the particle number density $n(r)$ and the inward streaming flow $v(r)$ are such that the particle flux entering the shell of radius $r$,  
\beq
\dot N = 4 \pi r^2 nv,
\label{eq:ndotdef}
\eeq 
is nearly independent of $r$ outside the region where the condensed baryons are accumulating. The flux of entropy into the shell  $r$ is $s\dot N$, where the entropy per particle is $s=k\ln T^{3/2}/n$, so the condition for a stationary energy distribution at radius $r$ is
\beq
kT\dot N{d\over dr}\ln {T^{3/2}\over n} =4\pi r^2\Lambda (T)n_e(n-n_e).
\label{eq:ndot}
\eeq
Expressed in terms of the cooling time (as in eq.~[\ref{eq:tauv}]) the particle number flux is
\beq
\dot N = {6\pi n r^3\over\tau}{d\ln r\over d\ln T^{3/2}/n}.
\label{eq:ndottau}
\eeq
Equations~(\ref{eq:ndotdef}) and~(\ref{eq:ndot}) are the steady-state versions of the mass and energy equations used in analyses of cluster cooling flows (Fabian 1988; Sarazin 1988).

Where the cooling time is greater than the free-fall time $\sim r/\sigma$ the plasma may be pressure-supported, and  one might use the condition that $\dot N$ is constant with the equation of hydrostatic equilibrium, 
\beq
{d\over dr}n_ekT = -{G M(<r)\over r^2}\mu n_em_p,
\eeq 
to compute the plasma temperature and density as functions of radius. We adopt a simpler approach, by noting that if the plasma temperature is independent of radius and the particle number density varies as $n\propto r^{-3/2}$ then equation~(\ref{eq:ndot}) indicates that $\dot N$ is constant, as wanted under the 
condition that significant amounts of mass are dropping out of the corona only near the optical radius of the galaxy.

\begin{figure}[t]
\plotone{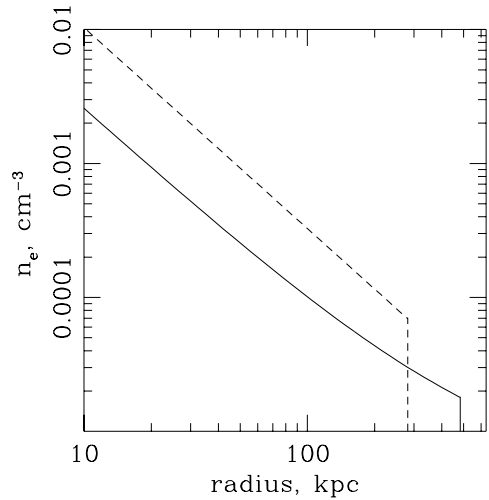}\caption{Electron number densities for Model~1 (the dashed line) and Model~2 (the solid line). Both take the plasma temperature to be $T=4T_v/3$. }\end{figure}

In corona Model 1 we take the mass distribution to be $\rho\propto r^{-2}$ (eq.~[\ref{eq:isosph}]) and we take the plasma temperature to be $T_e = 4T_v/3$ so the plasma density varies as $r^{-3/2}$, making $\dot N$ constant. This is a special case of Cavaliere's (1973) beta model with $\beta\equiv T_v/T_e= 3/4$. The dashed line in Figure~2 shows the electron number density in this model with the parameters in equations~(\ref{eq:sigma_late}) to~(\ref{eq:M_corona}). At heavy element abundance $10^{-1.5}$ times Solar and plasma temperature $T_e=4T_v/3=1.8\times 10^6$~K the Sutherland-Dopita cooling function is $\Lambda=10^{-23.02}$ erg~cm$^{3}$~s$^{-1}$. At primeval element abundance the cooling function is half this value. The mass flux $dM_{\rm cond}/dt = \mu m_p\dot N$ in this model for the corona, computed from equation~(\ref{eq:ndot}), is
\beqa
&&dM_{\rm cond}/dt = 14M_\odot \hbox{ yr}^{-1}\, (10^{-1.5}\hbox{ Solar)},\nonumber\\
&&dM_{\rm cond}/dt = 7M_\odot \hbox{ yr}^{-1} \hbox{(primeval)}.
\label{eq:mass_fluxes}
\eeqa
This is the rate at which mass is entering the condensed baryon components. It may be compared to the FP estimate of the present mean star formation rate per spiral galaxy (from the ratio of FP eqs. [30] and [52] with the spiral fraction  0.7),
\beq
dM_{\rm star}/dt = 1.7M_\odot\hbox{ y}^{-1}.
\label{eq:presentSFR}
\eeq
This is less than the flux of mass out of the corona in Model ~1 even at primeval element abundances. Since we know of no place where the condensed baryons might be accumulating at the rates in equation~(\ref{eq:mass_fluxes}) the indication is that $\dot N$  in Model~1 is too large by a  modest but significant factor. 

\begin{figure}[t]
\plotone{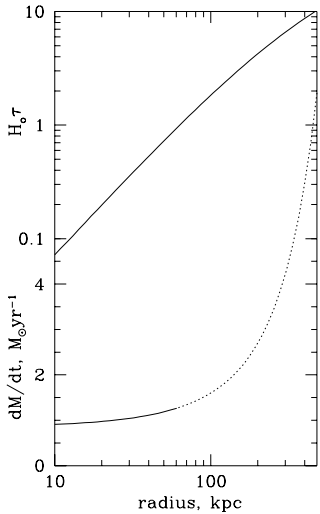}
\caption{The flux of mass out of the corona in Model~2 with heavy element abundance $10^{-1.5}$ times Solar. The upper curve is the ratio of the plasma cooling time to the Hubble time. The lower curve is the measure of the mass flux in equation~(\ref{eq:ndot}). Where the curve is dotted the cooling time is longer than the Hubble time and the expression for the mass flux need not apply.}
\end{figure}

We can remedy the problem by reducing $n_e$. In Model 2 this is done by adopting the density profile in the modified halo model in equation~(\ref{eq:modisosph}) with $\eta = 1$ while keeping the plasma temperature at $T=4T_v/3$. Since the mass distribution is not isothermal this is not a $\beta$ model, but at $r\la r_v$ the mass density run is close enough to $\rho\sim r^{-2}$ for our purpose. This is illustrated in Figure~3: at $r\la 60$~kpc the cooling time is less than the Hubble time, so we want $\dot N$ to be independent of radius, and we see that this is reasonably well satisfied at mass flux
\beq
dM_{\rm cond}/dt\simeq 1M_\odot \hbox{ yr}^{-1}\hbox{ (Model~2)}\, \label{eq:model_2_mass_flux}
\eeq
for heavy element abundance $10^{-1.5}$ times Solar, and half this value at primeval element abundances. At larger radii $\dot N$ shows a substantial variation with radius but this need not matter because here the plasma cooling time is long and the steady flow condition need not apply. 

The mass flux out of the corona in Model~2 (eq~[\ref{eq:model_2_mass_flux}]) is satisfactorily close to  the present star formation rate (eq.~[\ref{eq:presentSFR}]). We continue the discussion of observational tests of the massive corona models in \S 3.4 and \S 4, after commenting on the nature of the plasma distribution in our models.

In both corona models the plasma is warmer than the effective temperature of the dark matter. This moves the baryons toward the outer parts of the corona, making the baryon mass fraction in the central parts smaller than the cosmic value. This effect is more conspicuous in Model 2, where the halo is more extended,
further lowering the electron density in the inner part of the halo by moving more baryons to the outskirts. The suppression of the baryon mass fraction in the inner parts of a massive halo is observed in clusters of galaxies, as shown by Allen et al. (2004) and references therein, and we assume the cause of the suppression may be the same for clusters and massive galaxies: entropy is added to the plasma prior to and during assembly of the system.

\begin{figure}[t]
\epsscale{1.15}
\plotone{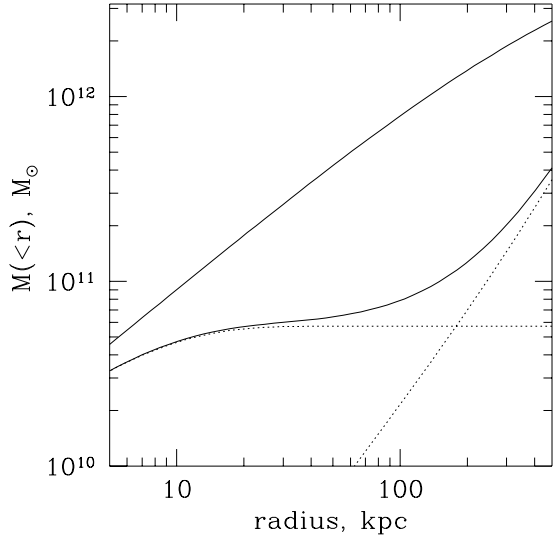}
\caption{The mass contained within a centered sphere in Model~2. The upper solid curve is the total included mass in baryons plus dark matter. The lower solid curve is the total baryon mass. The dotted curves show the two dominant baryon components, the plasma mass in the corona and the more concentrated baryon mass in condensed forms (eq.~[\ref{eq:condensed}]).}
\end{figure}

Figure~4 illustrates the situation in Model 2. The upper solid curve is the total mass contained within radius $r$ (eq.~[\ref{eq:modisosph}] with $\eta=1$) as a function of $r$. The lower solid curve is the sum of the corona and condensed baryon mass contained within $r$, and the dotted curves show the two separate components. We represent the condensed mass by the form 
\beq
M_{\rm cond}(<r) = {2\sigma_{\rm late}^2r_g\over G}
\left( 1 - e^{-r/r_g}\right).
\label{eq:condensed}
\eeq
There is no intended physical significance to this functional form: we choose it to satisfy the condition that the condensed baryon mass density is close to the total at small radius and we fix the parameter $r_g$ by the condition that the total condensed mass, $M_{\rm cond}=2\sigma_{\rm late}^2r_g/G$, is 13 percent of the total baryon mass in the corona (as discussed in \S 3.3). Under these conditions the radius that contains half of the condensed mass is $r_e=r_g \ln 2 = 4$~kpc. This is in the range of effective radii of $L\sim L_\ast$ spiral galaxies. Also, one sees in Figure~4 that in this representation the included condensed baryon mass at radius $r=10$~kpc is one half of the total, in agreement with what is known about the dark mass distribution in the Milky Way. Since the total condensed baryon mass agrees with the FP estimate from what is observed we conclude that this is a reasonable representation of the condensed mass distribution. 

At $r=20$~kpc the included baryon mass is close to the total condensed mass, $6\times 10^{10}M_\odot$, and the subdominant mass in the corona is causing the slow increase in included mass with increasing radius. The included masses in condensed baryons and the corona plasma are equal at $r=180$~kpc in Model~2. At  this radius the total baryon mass fraction is 0.09, or 56 percent of the cosmic value. This suppression of the baryon fraction is comparable to what is observed in relaxed clusters of galaxies at one quarter of the virial radius (Allen et al. 2004). The significant difference is that in spirals only half of the baryons within $r\sim 200$~kpc are in the coronal plasma, while the other half is in condensed forms. This difference might be expected from the shorter cooling times in galactic coronae. The included baryon mass fraction reaches the cosmic mean at radius $r_x$ (eq.~[\ref{eq:modisosph}]), at the right-hand edge of the figure. This would be replaced by a smooth transition to the cosmic value in a more realistic model for the corona. 
\subsection{Observational Tests}

Parameters characteristic of our two models are summarized in Table 1. We have already discussed the entry for the rate of settling of the corona; the results for Model~1 (eq.~[\ref{eq:mass_fluxes}]) motivate Model~2 (eq.~[\ref{eq:model_2_mass_flux}]). Here we consider other observational tests.

\begin{deluxetable}{lcc}
\tabletypesize{\scriptsize}
\tablecaption{Massive Corona Parameters\label{tbl-1}}
\tablewidth{0pt}
\tablehead{ \colhead{}  & \colhead{Model 1}  &  \colhead{Model 2} }
\startdata
Electron density $n_e$ at $r=10$~kpc, cm$^{-3}$ & 0.010 & 0.0026\\
$\bar n_e$ averaged from $r=10$ to 50~kpc, cm$^{-3}$ & 0.003 & 0.0007\\
$H_0\tau$ at the outer edge of the halo & 2.6 & 10.2\\
$\sigma_{\rm late}\tau/r$ at $r=10$~kpc & 4 & 14 \\
Mass flux $\mu\dot N$, $M_\odot$ y$^{-1}$  & 14 & 1\\
Compton-Thomson $y$ parameter & $10^{-7.0}$ & $10^{-7.5}$\\
$L_0(\epsilon >0.3\hbox{ keV},\rho>10\hbox{ kpc})$, erg s$^{-1}$
& $10^{40.6}$ & $10^{39.8}$\\
$L_0(\epsilon >0.3\hbox{ keV},10<\rho<20\hbox{ kpc})$, erg s$^{-1}$
& $10^{40.0}$ & $10^{38.8}$\\
Soft X-ray background $I_{\rm gc}$, $\hbox{ keV cm}^{-2}\hbox{ s}^{-1}\hbox{ sr}^{-1}\hbox{ keV}^{-1}$ & $\sim 1$ & $\sim 10$ \\
Central photoionization rate, s$^{-1}$ & $10^{-13.4}$ & $10^{-14.6}$\\
\enddata
\end{deluxetable}

\begin{figure}[t]
\epsscale{.9}
\plotone{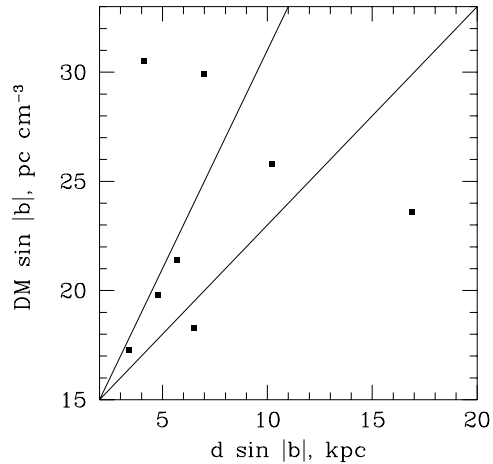}\caption{Dispersion measures along the lines of sight to pulsars in globular clusters. In order of increasing distance $d\sin |b|$ from the plane they are 47~Tuc, M15, M13, M5, M30, NGC1851, M3, and M53. The lines assume a close to homogeneous corona with electron density $n_e=0.001$~cm$^{-3}$ in the lower line and $n_e=0.002$~cm$^{-3}$ in the upper line.}
\end{figure}

If the Milky Way has a close to standard massive corona then the dispersion measures of pulsars outside the disk offer a useful probe of the plasma density in the inner corona. Manchester et al. (2005) list\footnote{We use the catalog in\\
 http://www.atnf.csiro.au/research/pulsar/psrcat} dispersion measures of pulsars in the Magellanic clouds and in eight globular clusters that are more than 2~kpc from the plane of the galaxy. Dispersion measures for the globular cluster pulsars are plotted in figure~5. The two lines assume a plane parallel plasma disk component with scale height $\ll 2$~kpc and surface density 15~pc~cm$^{-3}$ together with a near homogeneous corona. This value for the plasma surface density in the disk is not inconsistent with what is observed for pulsars closer to the central plane (Cordes \&\ Lazio 2005, fig. 1). There is not compelling evidence for detection of a corona outside the disk component, but one sees from the figure that the data would allow a corona with local number density in the range
\beq
0.001\hbox{ cm}^{-3}\la n_e\la  0.002\hbox{ cm}^{-3}.
\label{eq:nelocal}
\eeq 
The electron density in Model 2 at $r=10$~kpc, $n_e=0.0026$ cm$^{-3}$,  is reasonably consistent with this. In Model~1 $n_e$ is high, but we note in \S 4 that this could be because clumping of the plasma substantially enhances the cooling rate and lowers $n_e$  at $r\sim 10$~kpc.

Pulsars in the Magellanic Clouds have dispersion measures $\sim 100$ cm$^{-3}$~pc, which we can take as a bound on the contribution from the corona. This corresponds to mean density 
\beq
\bar n_e({\rm MC})\simeq  0.002\hbox{ cm}^{-3},
\label{eq:neMC}
\eeq
along the line of sight to the Magellanic Clouds. Table~1 lists the mean electron density averaged in the radial direction from $r=10$~kpc to $r=50$~kpc. The models are not inconsistent with equation~(\ref{eq:neMC}).  

Spitzer (1956) noted that the corona would be in pressure equilibrium with interstellar matter in the disk of the galaxy. Jenkins \& Tripp (2001) find that the thermal pressure in the disk of the Milky Way within a few kpc radius averages $nT=2000$~cm$^{-3}$~K, and they remark that this is not inconsistent with the weight of interstellar matter at the plane of the disk. In Model~2 the pressure at $r=10$~kpc is $nT=5000$~cm$^{-3}$~K. We do not know how to judge whether the difference is significant.

The integral of the electron density along a line of sight through the corona determines the effect of free electron scattering on the spectrum and anisotropy of the thermal 3~K background radiation (Rasmussen \& Pedersen 2001 and references therein). This is measured by the Compton-Thomson parameter $y = kT\sigma_{\rm T}\Sigma_e/(m_ec^2)$. For the numbers listed in Table~1 we compute the surface density $\Sigma_e$ of electrons integrated radially from 10~kpc to the outer edge of the corona, ignoring neighboring coronae. The empirical bound from the possible departure from a thermal spectrum is $y< 1.5\times 10^{-5}$ (Fixsen et al.1996). The variation $\delta y$ of this parameter across the sky gives the contribution to the radiation anisotropy, $\delta y\sim\delta T/T\ll 10^{-5}$, to avoid spoiling the consistency of the theory and observation of the anisotropy (Bennett et al. 2003). In both models $y$ is well below these constraints. The same is true of the contribution of  coronae to the mean electron pressure integrated across the Hubble length. 

The coronae we are considering are so large that the soft X-ray observation of a nearby galaxy constrains the X-ray surface brightness rather than the total luminosity of the galaxy. In the usual representation in terms of an effective luminosity, the integral of the surface brightness over the observed part of the galaxy is a recieved energy flux density $f$ and the effective luminosity is $L_0=4\pi D^2f$ if the distance to the galaxy is $D$. We compute this luminosity derived from a measurement of the energy flux density received from a cylinder of radius $\rho$ centered on the galaxy as
\beq
L_0 = 4\pi\int _0^{r_x}r^2dr F(r)j(r), 
\label{eq:Lx_exp}
\eeq
where $F(r)$ is the fraction of a sphere of radius $r$ centered on the galaxy that falls within the cylinder of radius $\rho$. Thus $F=1$ at $r<\rho$ and 
\beq
F(r) = 1-\sqrt{1-\rho ^2/r^2} 
\eeq
at larger radii. We approximate the plasma X-ray luminosity density at energy greater than $\epsilon$ as (Spitzer 1978)
\beq
j = 1.7\times 10^{-27} n_e^2T^{1/2}e^{-\epsilon/kT}
\hbox{ erg cm}^{-3}\hbox{ s}^{-1},
\label{eq:j(r)}
\eeq
in cgs units with the plasma temperature in $^\circ$K. 

The X-ray luminosities in Table~1 are computed at energy $\epsilon> 0.3$~keV, to match the measurements by Strickland et al. (2004a), and for the two models of the corona. The first of the entries refers to the energy flux density integrated from projected distance $\rho=10$~kpc to $\rho=r_x$ at the outer edge of the corona. The second entry refers to the energy flux density from the range of projected distances $10<\rho<20$~kpc, which is closer to available observations. There are several reasons to exclude the part at $\rho<10$~kpc: neutral hydrogen strongly absorbs soft X-rays, the corona plasma likely is strongly perturbed by the disk, and X-ray sources in the disk are not readily separated from what might be produced by the corona. 

Strickland et al. (2004a) find that the halo regions of the nearby galaxies NGC~891 and NGC~253, both of which have circular velocities similar similar to the Milky Way, have luminosities $\sim 10^{39}$~erg~s$^{-1}$ at 0.3 to 2~keV. This is  comparable to the luminosity at $10<\rho<20$~kpc in Model~2 and an order of magnitude below Model~1. The significance is unclear, however, because what is observed may have been produced by matter blown out of the disks by present or past starbursts, rather than the effect of a corona (e.g. Pietsch et al. 2000; Strickland et al. 2004b), and one must bear in mind that the luminosity computed for  Model~1 may be an overestimate for the reason discussed in \S 4. 

The total corona X-ray luminosity is constrained by the possible contribution to the soft X-ray background, which Wu, Fabian, \& Nulsen (2001) put at
\beq
I\la 4\hbox{ keV cm}^{-2}\hbox{ s}^{-1}\hbox{ sr}^{-1}\hbox{ keV}^{-1} 
\eeq
at photon energy $\epsilon\sim 0.25$~keV, after subtraction of detected sources. The contribution to $I$ by coronae depends on their history, which is even more uncertain than their present state, so we use a crude approximation,
\beq
I_{\rm gc} \sim {f_sn_gcLt\over8\pi\epsilon}
\label{eq:X-backgground}
\eeq
The characteristic galaxy number density is $n_g$ (eq.~[\ref{eq:ng}]), the spiral fraction is $f_s=0.7$, we approximate the time integral as $\int L\, dt/(1+z)\simeq0.5Lt$, with $t=10$~Gyr and $L$ constant at the values in Table~1 for $\rho>10$~kpc, and the factor $\epsilon =0.3$~keV in the denominator approximates the width of the spectral energy distribution. This estimate, listed in the second line from the bottom of Table~1, ignores the shift of the photon energies out of the band of detection, so if $L$ were close to constant it would be a considerable overestimate. Wu, Fabian, \& Nulsen (2001) point out that the corona X-ray luminosity could have been substantially larger in the past, meaning equation~(\ref{eq:X-backgground}) could be an underestimate, but that depends on how the galaxies and their coronae formed and evolved. We  conclude that our picture for the present-day state of massive coronae in spiral galaxies does not conflict with the soft X-ray background measurement, within the considerable uncertainties. 

The X-ray emission from coronae contributes to the local ionizing radiation, as Maloney \& Bland-Hawthorn (1999) note in their discussion of plasma distributed through the Local Group. An observer situated near the center of a corona would find that the radiation energy density produced by the corona is roughly isotropic at 
\beq
u = {1\over c}\int _{10~{\rm kpc}}^{r_x}j(r) dr,
\eeq
where $j(r)$ is the luminosity density (eq.~[\ref{eq:j(r)}]). As before, we assign the lower limit $r=10$~kpc because the situation at smaller radii is likely to be dominated by the disk. With photoionization cross section $\sigma\propto\nu^{-3}$, and $\sigma(\nu_o)=10^{-17.1}$~cm$^2$ at the hydrogen photoionization threshold, the photoionization rate near the center of the corona is
\beq
\Gamma \simeq \sigma_oc u/(3kT), 
\eeq
The numerical results are listed in Table~1. The Penton et al. ( 2000) estimate of the rate of ionization by the intergalactic background radiation, 
$\Gamma = 10^{-13.6}\hbox{ s}^{-1}$, is comparable to what might be expected from the corona in Model~1.  In Model~2 it appears that massive coronae would not be significant local sources of ionizing radiation. 

\section{Discussion}

We have argued for the viability of the idea that about half of the baryons are in gravitationally bound coronae in the dark matter halos of galaxies. In our models the mass distribution is designed to fit the galaxy-mass cross correlation function (Fig.~1), under the assumption of near monolithic massive halos, and the coronae are assumed to  have relaxed to a steady cooling flow. Under these conditions the corona electron density run is not likely to be much outside the two curves in Figure~2: a lower density than in Model~2 would bring the density in the outer parts of the corona closer to the global mean than seems reasonable for matter that has relaxed to dynamical equilibrium, and a density greater than or comparable to Model~1 would cause substantial problems with the observations.  

A better massive halo model would replace the sharp truncation of the mass and baryon densities at $r=r_x$ with a physically realistic transition to the intergalactic medium, but what is relevant for the inner structure of a corona in a near steady state is the situation where the plasma cooling time is comparable to the Hubble time, at $r\sim 60$~kpc: the plasma with longer cooling times serves as a reservoir supplying the flow to smaller radii. Conditions at this transition to near steady flow likely depend on departures from spherical symmetry, as may be caused by tidal interactions with neighboring massive halos or ongoing accretion. One might also have to consider rotational support in the inner parts of the corona, and possibly magnetic field pressure. But such refinements might await clearer evidence that massive coronae really exist. 

Model~2 appears to be observationally acceptable. It should be noted, however, that this model depends on the assumption of a smoothly distributed plasma, which may not be accurate, because the plasma is thermally unstable to the growth of density fluctuations that decrease the cooling times and so reduce the electron density run from what is shown in Figure~1.\footnote{In the range of radii $10\la r\la 60$~kpc the corona cooling time is less than the Hubble time but larger than the free fall time. Here we may apply the quasi-static approximation in eq.~[\ref{eq:ndottau}], where the density scales in proportion to the cooling time. This scaling ends at free fall.} That may make the X-ray luminosity of the corona significantly smaller than our estimate. The thermal instability may be expected to have a lesser  effect on the flux of mass from the corona into condensed forms of baryons, because this is determined by conditions where the cooling time and the thermal instability growth time both are comparable to the Hubble time. 

Model 1 appears to be marginally unacceptable: the most serious problem is the large rate of production of condensed baryons (eq.~[\ref{eq:mass_fluxes}]). It is worth noting, however, that if galactic winds removed about half of the baryons in the corona, leaving the rest with the radial distribution in Model~1, it would reduce the mass flux by a factor of four, leaving a still substantial baryon mass in coronae and an acceptable and interesting present rate of production of condensed forms of baryons. 

The near coincidence between the flux of mass out of the corona (eq.~[\ref{eq:model_2_mass_flux}]) and the present star formation rate (eq.~[\ref{eq:presentSFR}]) in Model 2 --- or in Model~1 with half the plasma mass we have been assuming --- agrees with Larson's (1972) conjecture that gas is flowing into the interstellar medium at the rate the interstellar medium is being converted to stars. Inflow to the disk at about this rate figures in current discussions of the history of the thin disk of the Milky Way (as reviewed by Pagel 1997; Matteucci 2001). The high velocity clouds often are mentioned as a source for the inflow, but the absence of analogs around other galaxies makes this unlikely (Pisano et al. 2004). Mass settling out of the corona seems to be a viable alternative. In this picture star formation in an isolated spiral galaxy will continue for a considerable time to come. 

We have not analyzed compatibility of the massive corona picture with the much cooler clouds of atomic hydrogen and low ionization ions detected near galaxies by quasar absorption line observations. Perhaps the plasma thermal instability produces the cloudy multiphase structure indicated by quasar absorption line studies (Churchill et al. 2000 and references therein). Chen et al. (2001) find that the HI surface density at projected distance $\rho=100$~kpc from an $L_\ast$ galaxy at modest redshift is typically $\sim 10^{17}$~cm$^{-2}$. Since the surface density of free electrons at $\rho=100$~kpc is $10^{20.1}$~cm$^{-2}$ in Model 2 there would be enough mass to produce the observed Ly$\alpha$ absorption. The minimum size of a region dense and cool enough to allow significant densities of HI and low ionization ions likely is set by thermal conduction by the motions of hydrogen atoms between formation and ionization. But we have left for future work the subtle analysis of the surface densities of these cool clouds. 

Another subject for more careful discussion is the history of the flux of mass out of the coronae into condensed forms of baryons. The star formation rate has to have been larger in the past, as has been widely discussed: an example of the evidence is the difference between the present star formation rate in equation~(\ref{eq:presentSFR}) and the mean rate of production of the observed present mass in stars and interstellar matter averaged over the age of the universe,
\beq
\langle dM_{\rm star}/dt\rangle = 
{\Omega_{\rm cold}\rho_{\rm crit}\over n_gt_0} = 6M_\odot\hbox{ y}^{-1},
\label{eq:meanSFR}
\eeq
where $\Omega_{\rm cold}=0.0035$ is the FP estimate of the condensed baryon mass and $t_0=14$~Gyr. Within the massive corona picture it is easy to imagine why the flux of mass out of the corona and into stars might be decreasing. We have estimated that at redshift $z=3$ the cooling time in the virialized mass concentration around a galaxy is less than the expansion time (eq.~[\ref{eq:Htau}]). That is, a corona assembled then would collapse to condensed matter in the disk or bulge. At lower redshift hierarchical growth of a galaxy would produce density fluctuations that would tend to increase the cooling rate and hence increase the flux of mass out of a corona present then. Settling of the denser clumps of plasma would tend to leave behind a smoother longer-lived corona. And at low redshifts supernovae may add entropy to the corona, further suppressing the rate of dissipative settling. 

We showed in \S 2 that the massive corona picture requires a close to smooth plasma distribution, not seriously disturbed by merging and accretion at low redshifts. There is empirical support for this condition, from the evidence that the Milky Way has not suffered a ``major invasion" since redshift unity (Gilmore, Wyse \&\ Norris 2001), and from the absence of the HI clouds around galaxies (Pisano et al. 2004) that would seem to be a natural consequence of merging and accretion at low redshifts (Blitz et al. 1999). 

The condition that we can neglect the effects of merging and accretion is not suggested by numerical simulations of the $\Lambda$CDM cosmology (as illustrated in Fig.~2 in Gao et al. 2004). It  may be important therefore that a possible remedy is under discussion: a long-range scalar interaction in the dark sector would hasten the assembly of dark matter halos and thus suppress the rate of major halo mergers at $z<1$ (Nusser et al. 2005 and references therein). Since baryons cannot significantly couple to the scalar force in this picture they would tend to accumulate in halos more slowly than the dark matter. This suggests yet another picture for the disposition of the baryons: maybe a substantial fraction have not fallen into massive halos. Or perhaps the galaxies did collect the full baryon mass fraction, but in a gentle way that allowed the formation of long-lived massive coronae. 

This paper is motivated by a challenging issue --- where are the baryons that are expected to have accompanied the dark matter in the formation of a galaxy? --- and by a natural idea --- that most of these baryons are in pressure-supported plasma, the  coronae, with radii comparable to that of the dark matter halos of $L\sim L_\ast$ galaxies. We conclude that the idea is viable but needs work. On the observational side, we will be following with particular interest the development of measurements of the soft X-ray surface brightnesses of isolated spiral galaxies at projected distances in the range of 100 to 300~kpc from the galaxy, which we expect would sample the bulk of a corona and, at the natural corona temperature in the range of one to a few times $10^6$~K, would be detectable if present. On the theoretical side, pressing issues are the compatibility of the corona picture with the cooler clouds identified by quasar absorption line studies and compatibility with the theory of structure formation within some acceptable version of the standard cosmology. 

\acknowledgments
We have benefitted from discussions with Charles Alcock, Avi Loeb, Ramesh Narayan, David Nice and David Strickland. MF received support for this work from the Monell Foundation at the Princeton Institute for Advanced Study and a Grant in Aid from the Japanese Ministry of Education at the University of Tokyo.

 \end{document}